\documentclass[aps,prd,twocolumn,10pt,groupedaddress]{revtex4-1}
\usepackage{amssymb}
\usepackage{graphicx}
\usepackage{amsmath}
\usepackage{hyperref}

\begin{document}

\title{The effect of Dilaton on holographic complexity growth}
\author{Yu-Sen An}
\email{anyusen@itp.ac.cn}
\author{Rong-Hui Peng}
\email{prh@itp.ac.cn}

\affiliation{CAS Key Laboratory of Theoretical Physics, Institute of Theoretical Physics, Chinese Academy of Sciences, Beijing 100190, China}
\affiliation{School of Physical Sciences, University of Chinese Academy of Sciences, No.19A Yuquan Road, Beijing 100049, P.R. China}
\date{\today}

\begin{abstract}
  In this paper, we investigate the action growth in various backgrounds in Einstein-Maxwell-Dilaton theory. We calculate the full time evolution of action growth in AdS dilaton black hole and find it approaches the late time bound from above. We investigate the black hole which is asymptotically Lifshitz and obtain its late-time and full time behavior. We find the violation of Lloyd bound in late time limit and show the full time behavior approaching the late time bound from above and exhibiting some new features for z sufficiently large. 
\end{abstract}
\maketitle

\section{Introduction}
The AdS/CFT correspondence shows that gravity theory in anti-de Sitter space is dual to conformal field theory on its boundary which implies a striking relation between the gauge theory and gravity theory~\cite{Maldacena:1997re}. As it is a weak-strong duality,  we are able to investigate various strong coupling effects in field theory from gravity side. A systematic rule of correspondence between two sides is proposed which is the so-called holographic dictionary~\cite{Gubser:1998bc,Witten:1998qj}. In 2006, Ryu and Takayanagi find that entanglement entropy in boundary field theory is dual to the minimal area of boundary anchored surface in the bulk~\cite{Ryu:2006bv, Ryu:2006ef}. This work illuminates the relation between quantum information theory and gravity theory  and implies that we can investigate quantum gravity effect via quantum information approach. 

In recent years, there is a growing interest in another field theory quantity called circuit complexity. In quantum computation theory,circuit complexity is defined to be the minimal number of simple gates required to build a typical state from a reference state within small tolerance $\epsilon$.   In order to inteprete  the growth of the size of Einstein-Rosen Bridge behind the horizon,  The authors in Ref~\cite{Susskind:2014rva}proposed that the  ERB size is dual to the quantum complexity on the boundary, which is called CV duality. Later, there is another  proposal which says that complexity is dual to on-shell action of the WdW patch~\cite{Brown:2015bva,Brown:2015lvg}, which is called CA duality. By calculating the action growth~\cite{Brown:2015bva,Brown:2015lvg,Lehner:2016vdi} , they find that the action obeys a bound called Lloyd bound~ \cite{Lloyd:2000} which is obtained in quantum computation using energy-time uncertainty principle. There are many works concerning this bound in various backgrounds \cite{Cai:2016xho, Cai:2017sjv,Swingle:2017zcd}. Moreover recent study \cite{HosseiniMansoori:2017tsm}shows the connection between the butterfly velocity and complexity growth rate. 

In the context of the string theory, as low energy limit is done,  a scalar field called dilaton occurs in the action . Dilaton couples to other fields in various nontrivial ways,   as an example, \cite{Garfinkle:1990qj} investigate the case that dilaton couples to the Maxwell field acting like the coupling constant of Maxwell action. The appearence of dilaton field changes the spacetime structure drastically and gives us more interesting black hole solutions. \cite{Garfinkle:1990qj} gets a dilaton black hole solution in asymptotically flat spacetime and \cite{Gao:2004tu} generalize this to asymptotically dS and AdS spacetime by introducing a Liouvile-type potential. Much interest has been paid to investigate the thermodynamic behavior of these black holes\cite{Sheykhi:2009pf,Li:2017kkj},  but the detailed analysis of the complexity behavior is still yet to be done.  

Despite the success of AdS/CFT correspondence, there are also many investigations beyond AdS . In AdS, space and time coordinate scale isotropically, while in order to describe some condensed matter systems , an anisotropically scaling is needed. Ref. ~\cite{Kachru:2008yh,Taylor:2008tg,Tarrio:2011de,Dong:2012se} come up with many backgrounds which support the scaling. 
\begin{equation}
\begin{split}
 t \to \lambda^{z} t \\
 x \to \lambda x
\end{split}
\end{equation}
Among them, Ref.\cite{Taylor:2008tg} finds that the Einstein-Maxwell-Dilaton system has such non-trivial black hole solution.While there are many investigations considering the thermodynamic and hydrodynamic properties of this solution \cite{Liu:2014dva,Pang:2009ad} , its complexity behavior is still under research.

In this work, we use CA proposal and investigate the action growth of various black hole solutions in  Einstein-Maxwell-Dilaton theory. In Section 2, we investigate a dilaton black hole in AdS vacuum, while an electro-magnet field exists, the Penrose diagram resembles the Schwartzchild case. While Ref.\cite{Cai:2017sjv}investigate the late-time behavior of this black hole, we investigate the full-time evolution of it and show that it approaches the late-time bound from above as Ref.\cite{Carmi:2017jqz}. In section 3, we go beyond the AdS case and investigate the Lifshitz-type black hole\cite{Taylor:2008tg} , we calculate the on-shell action in WdW patch of this black hole and find that it violates the Lloyd bound even in the late time limit. In Section 4, we discuss our result and give some interpretation.

\section {Full time evolution of action in Charged Dilaton Black hole in AdS space}
\subsection {charged dilaton black hole} 
We consider the Einstein-Maxwell-Dilaton theory with the action
\begin{equation}
S=\frac{1}{16 \pi} \int d^{4}x \sqrt{-g}(R-2(\partial \phi)^{2}-V(\phi)-e^{-2\phi}F^{2})
\label{bulka}
\end{equation}
where the Liouville-type potential is 
\begin{equation}
V(\phi)=-\frac{4}{l^{2}}-\frac{1}{l^{2}}[e^{2(\phi-\phi_{0})}+e^{-2(\phi-\phi_{0})}]
\end{equation}
When $\phi=\phi_{0}$, it reproduces the usual cosmological constant term in AdS space. Varying the action, we can get the equation of motion which is 
\begin{equation}
R_{\mu\nu}=2\partial_{\mu} \phi \partial_{\nu} \phi+\frac{1}{2} g_{\mu\nu} V+2 e^{-2\phi}(F_{\mu\alpha} F^{\alpha}_{\nu}-\frac{1}{4} g_{\mu\nu} F^{2})
\end{equation}
\begin{equation}
\partial_{\mu}(\sqrt{-g} e^{-2\phi} F^{\mu\nu})=0
\end{equation}
\begin{equation}
\partial^{2}\phi=\frac{1}{4} \frac{dV}{d\phi}-\frac{1}{2}e^{-2\phi} F^{2}
\end{equation}
there exists a static spherically symmetric black hole solution
\begin{equation}
ds^{2}=-f(r) dt^{2}+\frac{dr^{2}}{f(r)}+r(r-2D) d \Omega^{2}
\end{equation}
where $f(r)=1-\frac{2M}{r}+\frac{r(r-2D)}{l^{2}}$
the electro-magetic field and dilaton $\phi$ can be obtained via the equation of motion $F_{tr}=\frac{Q e^{2\phi}}{r(r-2D)}$, $e^{2\phi}=e^{2\phi_{0}}(1-\frac{2D}{r})$, $\phi_{0}$is the integration constant. D is the conserved dilaton charge which is $D=\frac{Q^{2} e^{2\phi_{0}}}{2M}$ \\
Unlike the usual RN case, although the system has electro-magnet field, because of the introduction of the dilaton coupling, there is a new curvature singularity located at the $r=2D$(between the inner horizon and outer horizon). It seems natural because of the instability of the inner horizon in RN black hole. Therefore, the penrose diagram looks like the AdS-Schwartzchild black hole rather than AdS-RN black hole.  \cite{Cai:2017sjv}investigated the late time behavior of the complexity growth and the result is 
\begin{equation}
\frac{d C}{dt}=2M-\mu Q- D
\end{equation}
and  \cite{Cai:2017sjv}proposed that the existence of dilaton will slow down the rate of complexity growth. Recent studies \cite{Carmi:2017jqz}says that in AdS-Schwartzchild case , the bound proposed in \cite{Cai:2017sjv}is approached from above so is violated in the early time. In the next section, we will show that the early-time violation behavior also occurs in this situation.
\begin{figure}
\includegraphics[width=0.4\textwidth]{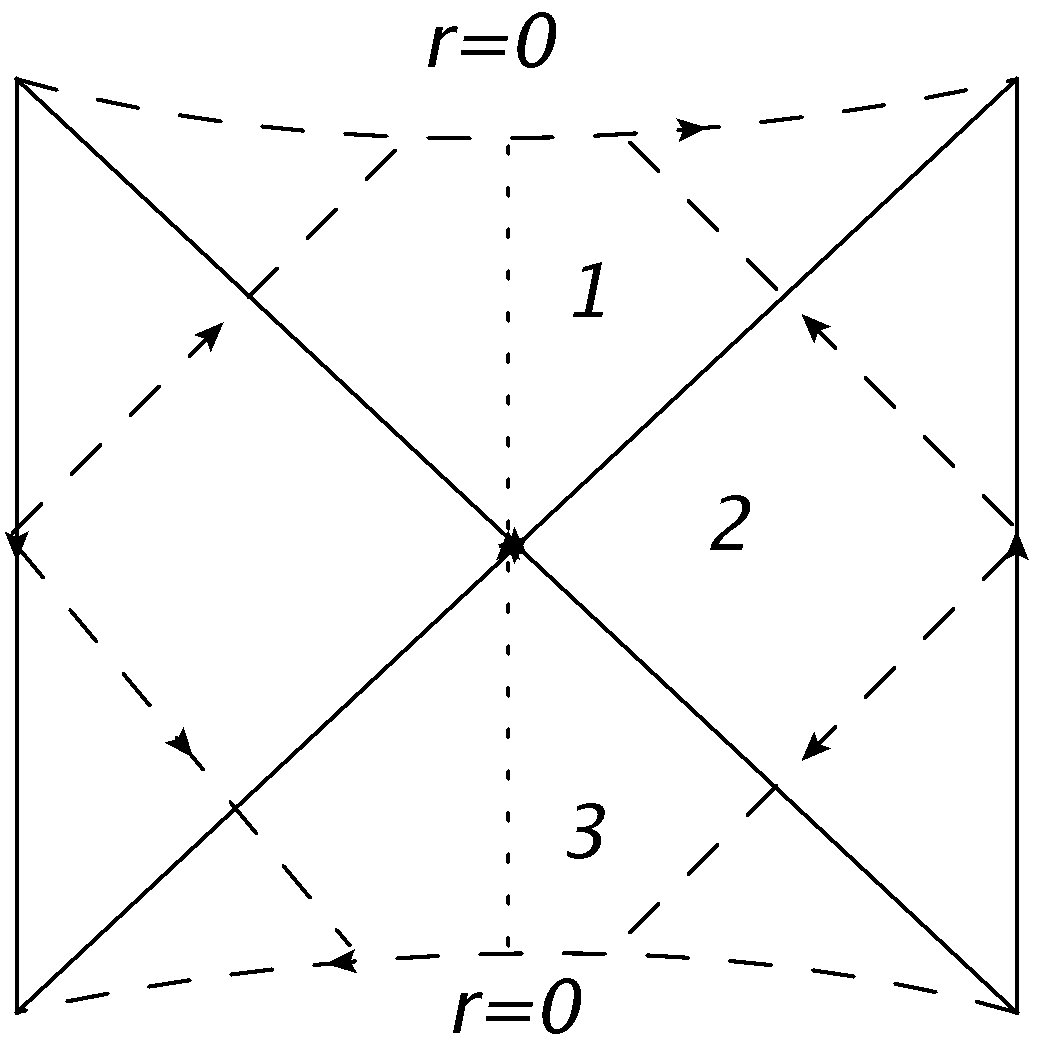}
\includegraphics[width=0.4\textwidth]{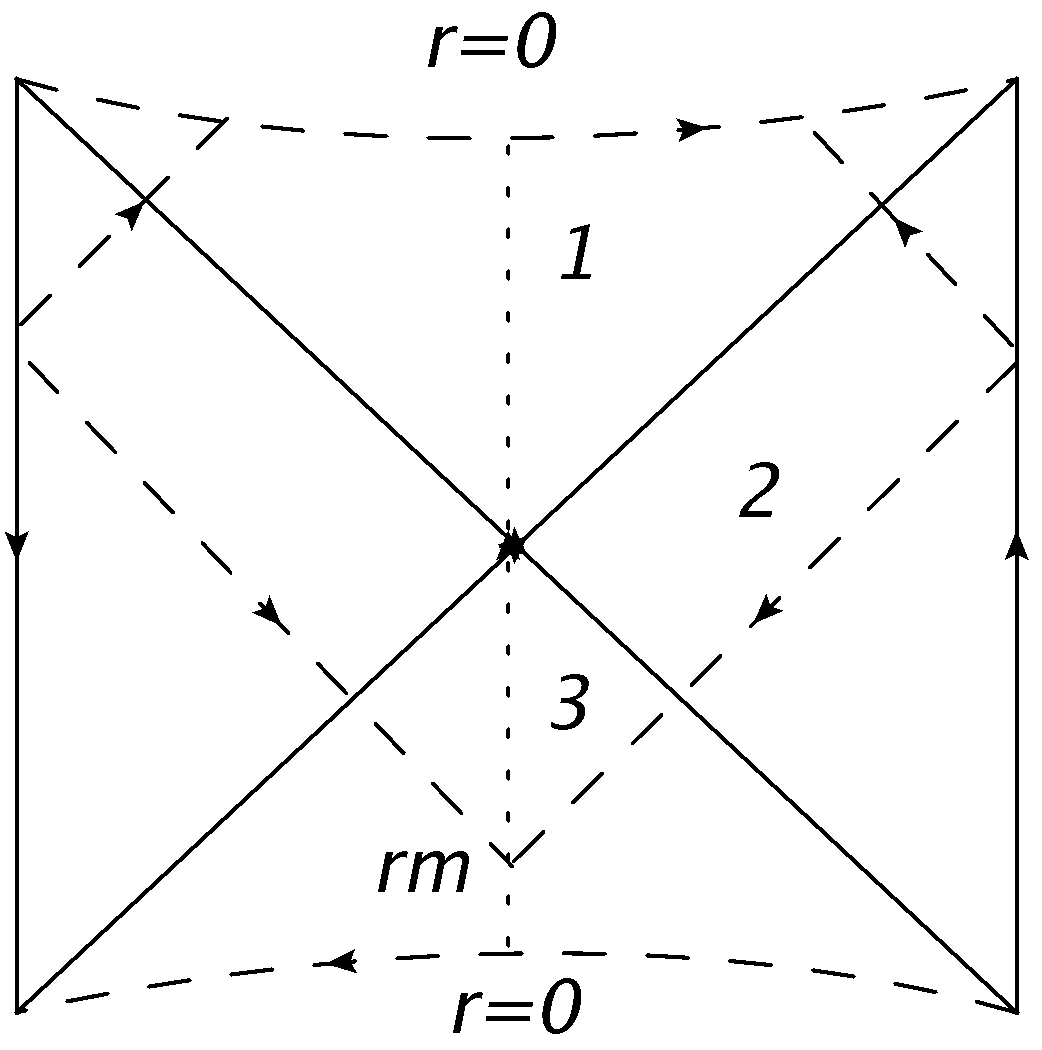}\\
  \caption{WdW patch of the time before(left side) and after (right side) the critical time, we assume boundary time satisfy the relation $t_{L}=t_{R}$, and at the right(left) boundary,  bulk time flows in the same(opposite) direction as the boundary, in calculating the bulk contribution of the total action ,we partition the spacetime into three regions }\label{fig:WDWpatch of the charged dilaton black hole}
\end{figure}

\subsection{General time dependence of the Action}
Ref \cite{Lehner:2016vdi}give a method to calculate the action in the presence of null boundary, the expression is as follows
\begin{equation}
\begin{split}
I = & \frac{1}{16 \pi G_N} \int_\mathcal{M} d^{4} x \sqrt{-g} \left(\mathcal R -2 \left(\partial \phi \right)^{2}-V (\phi)-e^{-2\phi}F^{2}\right) \\
&\quad+ \frac{1}{8\pi G_N} \int_{\mathcal{B}} d^3 x \sqrt{|h|} K + \frac{1}{8\pi G_N} \int_\Sigma d^{2}x \sqrt{\sigma} \eta
\\
&\quad -\frac{1}{8\pi G_N} \int_{\mathcal{B}'}
d\lambda\, d^{2} \theta \sqrt{\gamma} \kappa
+\frac{1}{8\pi G_N} \int_{\Sigma'} d^{2} x \sqrt{\sigma} a\,.
\end{split}
\end{equation}
Here, we follow the convention in \cite{Carmi:2017jqz,Carmi:2016wjl}. Terms in the expression above are respectively bulk term, GHY boundary term, Hayward joint term\cite{Hayward:1993my}, null boundary term and null joint term. We choose affine parametrization and set $ \kappa=0$ in the following, so the contribution of null boundary vanishes. 

We consider the full time evolution of action in this black hole(we set $G=1$ in this section for simplicity), we assume the boundary time have following relation $t_{L}=t_{R}=\frac{t}{2}$, the time dependence have two stage. First, the past null-boundary intersect the past-singularity, and we have the past GHY boundary term, after an amount of time ,which we call critical time, the two past-null boundary intersect each other , and a null joint term replace the GHY boundary term. The complexity behavior is different in the time below and above the critical time. So, we first determine the critical time $t_{c}$\\
\begin{equation}
\frac{t_{c}}{2}-r^{*}(\infty)=t-r^{*}(0)
\end{equation}
\begin{equation}
-\frac{t_{c}}{2}+r^{*}(\infty)=t+r^{*}(0)
\end{equation}
the critical time is 
\begin{equation}
t_{c}=2(r^{*}(\infty)-r^{*}(0))
\end{equation}
when $t< t_{c}$
the contribution contains three part
\begin{equation}
S=S_{bulk}+S_{GHY}+S_{joint}
\end{equation}
the bulk contribution is the Einstein-Hilbert action plus matter field action which is ($\ref{bulka}$), and the surface comes from past and future singularity surface and the cutoff surface at $r_{max}$. The joint term is null-spacelike /null-timelike joint which occurs at singularity / cutoff surface respectively.  
The bulk term is 
\begin{equation}
S=\frac{1}{16 \pi} \int d^{4}x [4r(r-2D)+(r-2D)^{2}+r^{2}-\frac{2Q^{2}l^{2} e^{2\phi_{0}}}{r^{2}}]
\end{equation}
It is convenient to divide the integral region into three portions and the result is as follows
\begin{equation}
\begin{aligned}
S_{1}=-\frac{1}{4l^{2}} \int _{2D}^{rh} dr [ &4r(r-2D)+(r-2D)^{2}+r^{2}\\
&-\frac{2Q^{2}l^{2} e^{2\phi_{0}}}{r^{2}}](\frac{t}{2}+r^{*}_{\infty}-r^{*}(r))
\end{aligned}
\end{equation}
\begin{equation}
\begin{aligned}
S_{2}=-\frac{1}{2l^{2}} \int _{rh}^{rmax} dr [& 4r(r-2D)+(r-2D)^{2}+r^{2}\\
&-\frac{2Q^{2}l^{2} e^{2\phi_{0}}}{r^{2}}](r^{*}_{\infty}-r^{*}(r))
\end{aligned}
\end{equation}
\begin{equation}
\begin{aligned}
S_{3}=-\frac{1}{4l^{2}} \int _{2D}^{rh} dr [ &4r(r-2D)+(r-2D)^{2}+r^{2}\\
&-\frac{2Q^{2}l^{2} e^{2\phi_{0}}}{r^{2}}](-\frac{t}{2}+r^{*}_{\infty}-r^{*}(r))
\end{aligned}
\end{equation}
We see the time dependence cancels each other and bulk term is independent of time. 
Then, we calculate the surface term at singularity. We choose 
\begin{equation}
n_{\alpha}=-| f(r)|^{-1/2} \partial_{\alpha}r
\end{equation}
and the extrinsic curvature is 
\begin{equation}
K=\nabla_{\alpha} n^{\alpha}=\frac{1}{U^{2}} \frac{d}{dr}(U^{2} n^{r})
\end{equation}
where $U^{2}=r(r-2D)$
So the GHY action is 
\begin{equation}
I_{future}=\frac{1}{2} |f|^{1/2} \frac{d}{dr}(r(r-2D)|f|^{1/2})(\frac{t}{2}+r^{*}_{\infty}-r^{*}(r))|_{r=2D}
\end{equation}
\begin{equation}
I_{past}=\frac{1}{2}|f|^{1/2} \frac{d}{dr}(r(r-2D)|f|^{1/2})  (-\frac{t}{2}+r^{*}_{\infty}-r^{*}(r))|_{r=2D}
\end{equation}
\begin{equation}
I_{cutoff}=|f|^{1/2} \frac{d}{dr}(r(r-2D) |f|^{1/2})(r^{*}_{\infty}-r^{*}(r)) |_{r=r_{max}}
\end{equation}
We see the cancelation between surface at the past and future singularity, and from \cite{Chapman:2016hwi} we know that the joint terms are independent of time. Combine the above result , we see the action is constant until $t=t_{c}$. \\

When $t>t_{c}$, null joint forms at $r=r_{m}$and there is no surface term from past singularity. $r_{m}$ is obtained by equation
\begin{equation}
\frac{t-t_{c}}{2}+r^{*}(r_{m})-r^{*}(0)=0
\end{equation}
The null joint term depends on time implicitly through $r_{m}$ \\
The bulk action is 
\begin{equation}
\begin{aligned}
I_{bulk}=I_{bulk}^{0}-\frac{1}{4L^{2}}\int _{2D}^{rm} dr [& 4r(r-2D)+(r-2D)^{2}+r^{2}\\
&-\frac{2Q^{2}l^{2} e^{2\phi_{0}}}{r^{2}}](\frac{t}{2}-r^{*}_{\infty}+r^{*}(r))
\end{aligned}
\end{equation}
So the change of the bulk action compared with the $t<t_{c}$ case is 
\begin{equation}
\begin{aligned}
\delta I_{bulk}=-\frac{1}{4L^{2}} \int _{2D}^{rm} dr [ &4r(r-2D)+(r-2D)^{2}+r^{2}\\
&-\frac{2Q^{2}l^{2} e^{2\phi_{0}}}{r^{2}}](\frac{\delta t}{2}+r^{*}(r)-r^{*}(0))
\label{bulka2}
\end{aligned}
\end{equation}
where $\delta t=t-t_{c}$
because of the lack of surface term of past singularity, the surface contribution also depends on t. 
\begin{equation}
I_{surf}=I_{0}-I_{past}
\end{equation}
so
\begin{equation}
\delta I_{surf}=\frac{1}{2}\frac{d}{dr}(r(r-2D)|f|^{1/2}) |f|^{1/2} (\frac{\delta t}{2}+r^{*}(r)-r^{*}(0))|_{r=2D}
\label{surf2}
\end{equation}
For null joint term
\begin{align}
k_{R}=-\alpha dt + \alpha \frac{dr}{f(r)} \\
k_{L}=\alpha dt + \alpha \frac{dr}{f(r)}
\end{align}
\begin{equation}
a=log(-\frac{1}{2}k_{R} \cdot k_{L})=-log(\frac{|f|}{\alpha^{2}})
\end{equation}
so the joint term reads
\begin{equation}
\begin{split}
&I_{jnt}=\frac{1}{8\pi } \int_{Sigma'} d^{2}x \sqrt{\sigma} a \\
&\quad =-\frac{r_{m}(r_{m}-2D)}{2} log\frac{|f(r_{m})|}{\alpha^{2}}\,.
\end{split}
\label{jnt2}
\end{equation}
combine the above result ($\ref{bulka2}$, $\ref{surf2}$,$\ref{jnt2}$) together  and take derivative with respect to t , recall that $\frac{dr_{m}}{dt}=-\frac{1}{2}f(r_{m})$, we finally get the action growth rate at time $t>t_{c}$
\begin{equation}
\frac{dI_{tot}}{dt}=2M-{\mu_{m}}Q-D+\frac{1}{2}(r_{m}-D)f(r_{m}) log \frac{|f(r_{m})|}{\alpha^{2}}
\end{equation}
where $ \mu_{m}=\frac{Qe^{2\phi_{0}}}{r_{m}}$
at late time limit $r_{m} \to r_{+}$, we see $\mu_{m}$becomes the chemical potential and the last term vanishes. So we recover the late time result in \cite{Cai:2017sjv}. The rate of growth of action for $t>t_{c}$ is plotted as Fig2. 
\begin{figure}
\includegraphics[width=0.45\textwidth]{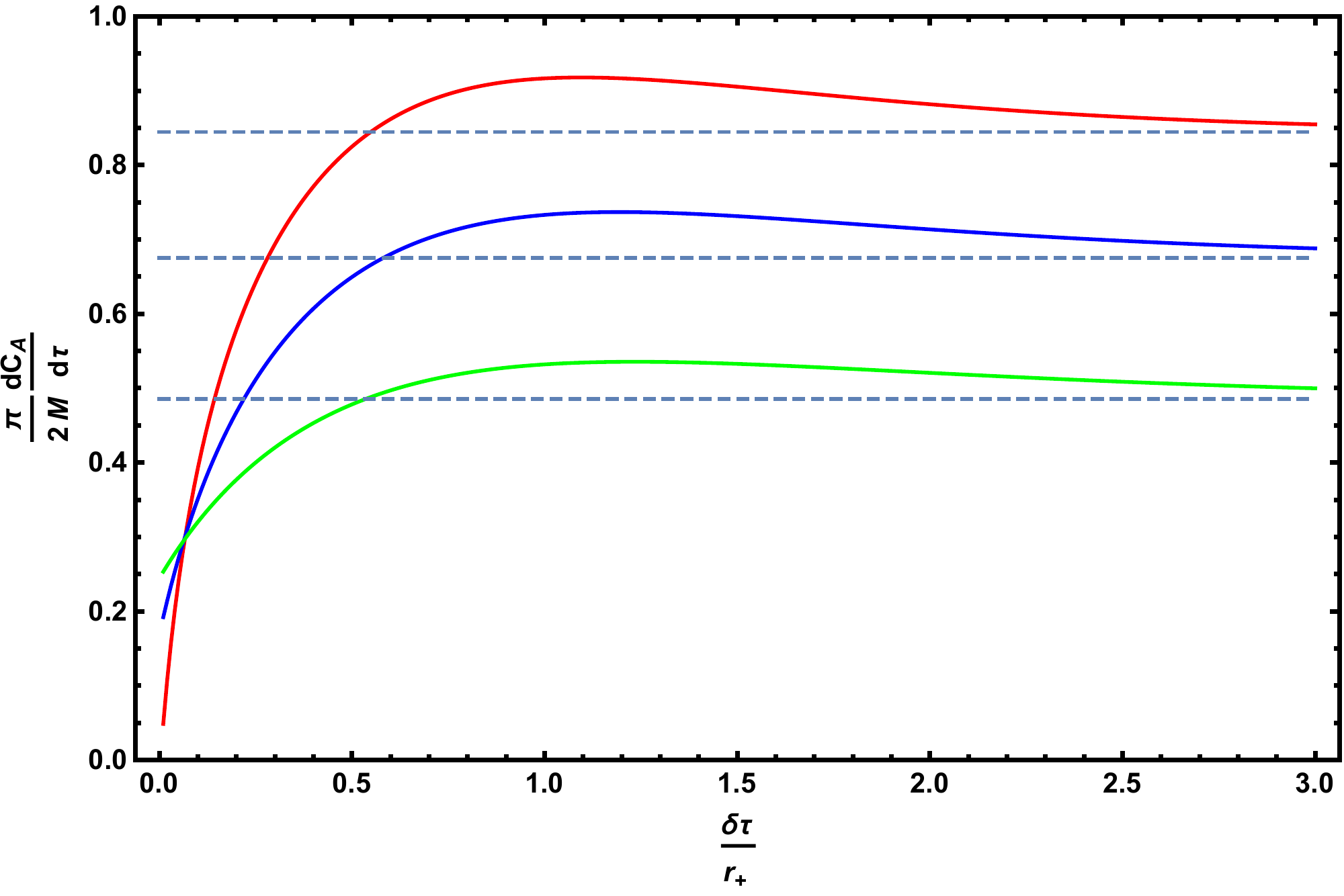}\\
  \caption{There are two parameters, $y=\frac{\mu Q}{2M}$, $z=r_{+}/L$. For this picture, we fix $z=1$, the green line correspond to $y=0.1$, blue line $y=0.2$, red line $y=0.3$. we find the complexity growth rate approaches the late time bound from above.}
  \label{fig:complexity growth rate in different chemical potential}
\end{figure}
\section{Action growth in Lifshitz-Like black brane}
\subsection{Lifshitz black brane in EMD theory}
In many condensed matter systems, near the critical point, anisotropic scaling symmetry is expected. The gravitational systems which have the same scaling behavior are constructed in various situations.Ref.\cite{Kachru:2008yh}adds high order form field,while Ref.\cite{Taylor:2008tg} realizes Lifshitz spacetime using massive vector field and Einstein-Maxwell-Dilaton theory. Here we choose the solution constructed in Ref.\cite{Taylor:2008tg}by coupling the dilaton field to an Abelian Maxwell field. The thermodynamic beheavior of Lifshitz spacetime is investigated in Ref.\cite{Liu:2014dva} 

We consider the following action in (d+2)-dimensional spacetime
\begin{equation}
\begin{aligned}
       I=\frac1{16\pi G_{d+2}}\int d^{d+2} x\sqrt{-g}[&R-2\Lambda-\frac12 \partial_\mu\phi\partial^\mu\phi \\
       &-\frac14 e^{\lambda\phi}F_{\mu\nu}F^{\mu\nu}].
\end{aligned}
\end{equation}
where $\Lambda\ $is the cosmological constant and the matter fields are a massless scalar and an abelian gauge field.We can get the equations of motion :
\begin{equation}
\label{2eq2}
\partial_{\mu}(\sqrt{-g}e^{\lambda\phi}F^{\mu\nu})=0,
\end{equation}
\begin{equation}
\partial_{\mu}(\sqrt{-g}\partial^{\mu}\phi)-\frac{\lambda}{4}\sqrt{-g}e^{\lambda\phi}F_{\mu\nu}F^{\mu\nu}=0,
\end{equation}
\begin{equation}
\label{2eq4} R_{\mu\nu}=\frac{2}{d}\Lambda
g_{\mu\nu}+\frac{1}{2}\partial_{\mu}\phi\partial_{\nu}\phi+\frac{1}{2}e^{\lambda\phi}F_{\mu\rho}{F_{\nu}}^{\rho}
-\frac{1}{4d}g_{\mu\nu}e^{\lambda\phi}F_{\mu\nu}F^{\mu\nu}.
\end{equation}
which has the following asymptotic Lifshitz black hole solution:
\begin{eqnarray}
\label{2eq11}
&ds^{2}=L^{2}(-r^{2z}f(r)dt^{2}+\frac{dr^{2}}{r^{2}f(r)}+r^{2}\sum\limits^{d}_{i=1}dx^{2}_{i}),~~~
\\
&f(r)=1-\frac{r^{z+d}_{+}}{r^{z+d}}\\
&F_{rt}=qe^{-\lambda\phi}r^{z-d-1},~~~e^{\lambda\phi}=r^{\lambda\sqrt{2(z-1)d}},\nonumber\\
&\lambda^{2}=\frac{2d}{z-1},~~~q^{2}=2L^{2}(z-1)(z+d),\nonumber\\
&\Lambda=-\frac{(z+d-1)(z+d)}{2L^{2}}.
\label{eom}
\end{eqnarray}

We can obtain the temperature via Euclidean path integral 
\begin{align}
 T_{H}=\frac{(z+d)r^{z}_{+}}{4\pi},
 \end{align}
 and the black hole entropy  
 \begin{equation}
 S_{BH}=\frac{\Omega_{d}L^{d}}{4G_{d+2}}(\frac{4\pi}{z+d})^{\frac{d}{z}}T^{\frac{d}{z}}
\end{equation}
where $\Omega_{d}$ denotes the volume of the $d$-dimensional spatial coordinates.

\subsection{late time behavior in Lifshitz black brane}
The volume contribution is
\begin{equation}
S_{V}=\int_{V}(R-2\Lambda-\frac12\partial_\mu\phi\partial^\mu\phi-\frac14e^{\lambda\phi}F_{\mu\nu}F^{\mu\nu})\sqrt{-g}d^{n+2}x
\end{equation}
We will use the null coordinates $u$ and $v$, defined by
\begin{equation}
du := dt + \frac1{r^{z+1} f} dr,~~~~
dv := dt - \frac1{r^{z+1} f} dr
\end{equation}
Integrating these relations yields the ``infalling" null coordinate $u
= t + r^*(r)$ and the ``outgoing" null coordinate $v = t - r^*(r)$,
where $r^*(r) := \int \frac1{r^{z+1} f} dr$. The metric becomes
\begin{equation}
ds^2 = L^2[-r^{2z}fdu^2 + 2r^{z-1}dudr + r^2 \sum\limits^{d}_{i=1}dx^{2}_{i}]
\end{equation}
or                                                                                                                                                                                                                              
\begin{equation}
ds^2 =L^2[-r^{2z}f dv^2 - 2r^{z-1}dvdr + r^2 \sum\limits^{d}_{i=1}dx^{2}_{i}]
\end{equation}
when expressed in terms of the null coordinates. For the three choices
$(t,r)$, $(u,r)$, and $(v,r)$ we have 
\begin{equation}
\int \sqrt{-g} d^{d+2} x = \Omega_{d}L^{d+2}\int r^{d+z-1} dr dw,
\end{equation}
where $w = \{ t, u, v \}$.

\begin{figure}
\includegraphics[width=0.55\textwidth]{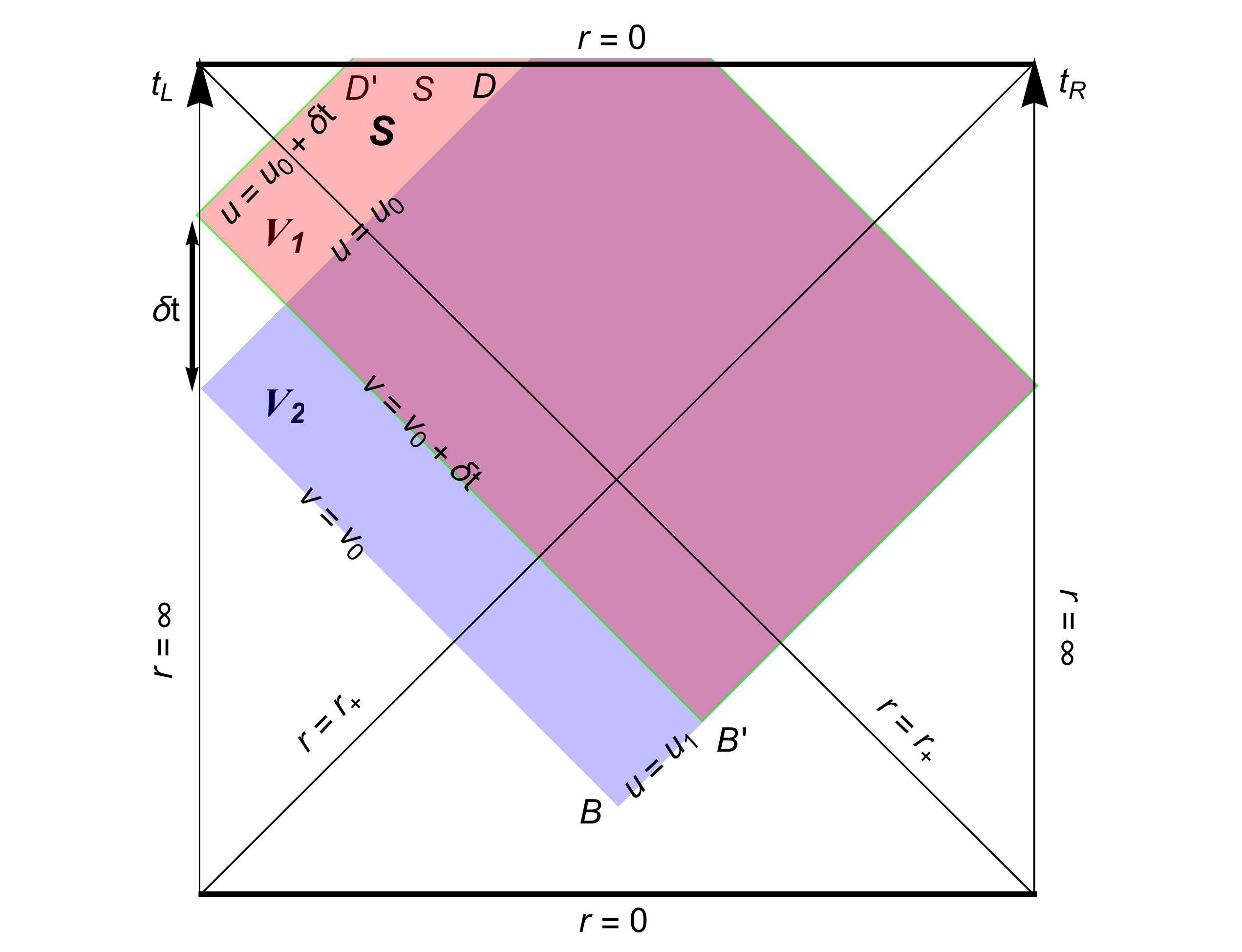}\\
  \caption{WdW patch of Lifshitz black hole}
\end{figure}

As shown in Fig.3,the region $V_1$ is bounded by the null surfaces $u=u_0$,$u=u_0+\delta t$,$v=v_0+\delta t$,$r=\epsilon$.The volume integral is best performed in the $(u,r)$ coordinate system, in this system the surface $v=v_0+\delta t$ is described by $r=\rho_0(u)$,with $r^*(\rho_0)=\frac12(v_0+\delta t-u)$.Making use of equations of motion,we get 
\begin{equation}
\begin{aligned}
R-2\Lambda-\frac12 \partial_\mu\phi\partial^\mu\phi-\frac14 e^{\lambda\phi}F_{\mu\nu}F^{\mu\nu}&=\frac{4\Lambda}{d}-\frac1{2d} e^{\lambda\phi}F_{\mu\nu}F^{\mu\nu} \\
&=\frac{-2(z+d)}{L^2}
\end{aligned}
\end{equation}
with this we have 
\begin{equation}
\begin{aligned}
S_{V_1} &= -2(d+z)L^d\Omega_d \int^{u_0+ \delta t}_{u_0} du
\int_{\epsilon}^{{\rho_0}(u)} r^{d+z-1}dr\\
&= -2L^d\Omega_d \int^{u_0+ \delta t}_{u_0} du
[{\rho_0}^{d+z}(u)]
\end{aligned}
\end{equation}
Where we neglect the $\epsilon^{n+1}$term in the integral as $\epsilon \to 0$
. The region $V_2$ is bounded by the null surfaces $u=u_0$, $u=u_1$,
$v=v_0$, and $v=v_0+\delta t$. In this case, the volume integral is most easily
performed in the $(v,r)$ coordinates, in which the surfaces $u=u_{0,1}$ are
described by $r=\rho_{0,1}(v)$, with $r^*(\rho_{0,1}) = \frac{1}{2}(v-u_{0,1})$. Then we have
\begin{equation}
\begin{aligned}
S_{V_2} &= -2(d+z)L^d\Omega_d \int^{v_0+ \delta t}_{v_0} dv
\int_{{\rho_1}(v)}^{{\rho_0}(v)} r^{d+z-1}dr\\
&= -2L^d\Omega_d \int^{v_0+ \delta t}_{v_0} dv
[{\rho_0}^{d+z}(v)-{\rho_1}^{d+z}(v)].
\end{aligned}
\end{equation}
Using a variables change :$u = u_0+v_0 + \delta t - v$ , we can see the terms involving ${\rho_0}(u)$ and
${\rho}_0(v)$ cancel out. We are left with
\begin{equation}
S_{V_1} - S_{V_2} = -2L^d\Omega_d \int^{v_0+ \delta t}_{v_0} dv
[{\rho_1}^{d+z}(v)].
\end{equation}
with the function $\rho_1(v)$ varying from $r_{B}$ to $r_{B}+O(\delta t)$,and hence the volume contribution to $\delta S$ is simply
\begin{equation}
\begin{aligned}
S_{V_1} - S_{V_2} =-2L^d\Omega_d {r_B}^{d+z}\delta t
\label{dS:volume}
\end{aligned}
\end{equation}
The surface contribution to $\delta S$ are given by
$-2 \int_{S} K\, d\Sigma$, where $S$ is the boundary segment given by the spacelike
hypersurface $r = \epsilon$. The (future-directed) unit normal is given by
$n_\alpha = \frac{L}{r}|f|^{-1/2} \partial_\alpha r$. The extrinsic curvature is   then
\begin{equation}
K = \nabla_\alpha n^\alpha = -\frac{1}{L^{d+2} r^{z+d-1}} \frac{d}{dr}
\Bigl(L^{d+1} r^{z+d} |f|^{1/2} \Bigr),
\end{equation}
and the volume element:
\begin{equation}
d\Sigma = \Omega_{d} L^{d+1}|f|^{1/2} r^{z+d} dt
\end{equation}
Letting $r = \epsilon
\ll r_{+}$ and then approximating $f \simeq -(r_{+}/r)^{z+d}$;  $K\simeq -\frac{z+d}{2L}(r_{+}/r)^{\frac{z+d}2}$; $d\Sigma \simeq \Omega_{d} L^{d+1}(r_+r)^{\frac{z+d}2} dt$,
we find that
\begin{equation}
-2 \int_{S} K d\Sigma = (z+d) \Omega_{d}L^{d}{r_+}^{z+d} \delta t.
\label{dS:surface}
\end{equation}
It is finite and independent of $\epsilon$. 

We then calculate joint terms at $ B$,$B'$.The null joint rule states that \cite{Lehner:2016vdi}
\begin{equation}
a = \ln\bigl( -{\textstyle \frac{1}{2}} k \cdot \bar{k} \bigr),
\end{equation}
We choose the vectors $k^\alpha$ and $\bar{k}^\alpha$ to be
affinely parametrized and read
\begin{equation}
\begin{aligned}
k_\alpha = -c\partial_\alpha v = -c\partial_\alpha(t - r^*), \qquad
\bar{k}_\alpha= \bar{c}\partial_\alpha u
= \bar{c}\partial_\alpha(t+r^*)\,
\end{aligned}
\end{equation}
where $c$ and $\bar{c}$ are arbitrary (positive) constants.With these
choices, we have that $k \cdot \bar{k} = 2c\bar{c}/(f L^2 r^{2z})$, then
\begin{equation}
a = \ln\bigl[-c\bar{c}/(f L^2 r^{2z})]£¬
\end{equation}
 With the above expression, we find that
\begin{align}
2 \oint_{B'} a\, dS - 2 \oint_{B} a\, dS
= 2\Omega_d \bigl[ h(r_{B'}) - h(r_{B})\bigr],
\end{align}
where $h(r) := r^d L^d \ln[-c\bar{c}/(f L^2 r^{2z})]$.

Then we perform a Taylor expansion of
$h(r)$ about $r = r_B$. Because the displacement is in a direction of
increasing $v$, we have that $du = 0$, $dv = \delta t$, and
$dr = -\frac{1}{2} f r^{z+1}\delta t$. This gives us
\begin{equation}
\begin{aligned}
&h(r_{B'}) - h(r_{B}) = -\frac{1}{2} f r^{z+1}\frac{dh}{dr}\bigg|_{r=r_{B}} \delta t\\
&= -\frac{1}{2} L^d r^{z+d}\biggl[ -r\frac{df}{dr} -2z f+ fd
\ln\biggl(\frac{c\bar{c}}{-fL^2 r^{2z}} \biggr) \biggr]\bigg|_{r=r_{B}} \delta t,
\end{aligned}
\end{equation}

and then
\begin{equation}
\begin{aligned}
&2 \oint_{B'} a dS - 2 \oint_{B} a dS \\
&= -\Omega_{d} L^d r^{z+d}\biggl[ -r\frac{df}{dr} -2z f+ fd
\ln\biggl(\frac{c\bar{c}}{-fL^2 r^{2z}} \biggr) \biggr]\bigg|_{r=r_{B}} \delta t
\end{aligned}
\end{equation}

Making use of the explicit expression of
$f$,and take the late time limit $ r_{B} \to r_{+}$,the Log term vanishes,and the result is 
\begin{equation}
2 \oint_{B'} a dS - 2 \oint_{B} a dS
=(z+d)L^d\Omega_d {r_{+}}^{d+z}\delta t
\label{dS:joint}
\end{equation}

Combining Eqs.~(\ref{dS:volume}), (\ref{dS:surface}), and
(\ref{dS:joint}), we have
\begin{equation}
\delta S = (2z+2d-2)L^d\Omega_d {r_+}^{d+z}\delta t
\end{equation}
in late time limit.
 From reference \cite{Pang:2009ad},we know the mass of the Lifshitz spacetime, this is
 \begin{equation}
 M=\frac{r^{z+d}_{+}dL^{d}\Omega_{d}}{16\pi G_{d+2}},
 \label{m}
\end{equation}
This mass is the Komar integral subtracting the zero temperature background to remove the infinite volume divergence\cite{Taylor:2008tg}. So we have,
\begin{equation}
\frac{dS}{dt} = 32\pi G_{d+2}\frac{z+d-1}{d}M
\end{equation}
and in a more usual convention,
\begin{equation}
\frac{dI}{dt} = 2\frac{z+d-1}{d}M
\end{equation}
we can see that,when $z=1$,
\begin{equation}
\frac{dI}{dt} = 2M
\end{equation}
it is just the result in reference \cite{Lehner:2016vdi}.
When $z>1$,we have $\frac{dI}{dt} >2M$, the bound in \cite{Brown:2015bva,Brown:2015lvg} is violated even in the late time limt.  

\subsection{general time behavior Lifshitz black brane}
In this subsection, we investigate the full time evolution of action growth and see whether the late time result is approached from below or above.We divide the Wheeler-DeWitt patch into three parts as in Fig1. When $t<t_c$,the action is composed of  following three parts.
\begin{equation}
I_{tot}=I_{bulk}+I_{GHY}+I_{joint}
\end{equation}
where $I_{bulk}$ is the bulk terms  and $I_{GHY}$ is the GHY surface terms at $r=\epsilon\to 0$ and$r=r_{max}\to \infty $ ,the joint terms is the joint formed by null and spacelike(timelike) surface at $r=\epsilon$ ($r=r_{max}$) .
For $t<t_c$,we have
\begin{equation}
I_{bulk}^1=-\int^{r_+}_{\epsilon_0}\frac{(z+d)L^d\Omega_d}{8\pi G}(\frac{t}2+r^*(\infty)-r^*(r))r^{z+d-1}dr
\end{equation}
\begin{equation}
I_{bulk}^2=-\int^{r_{max}}_{r_+}\frac{(z+d)L^d\Omega_d}{8\pi G}2(r^*(\infty)-r^*(r))r^{z+d-1}dr
\end{equation}
\begin{equation}
I_{bulk}^3=-\int^{r_+}_{\epsilon_0}\frac{(z+d)L^d\Omega_d}{8\pi G}(-\frac{t}2+r^*(\infty)-r^*(r))r^{z+d-1}dr
\end{equation}
we see total cancelation and the bulk term is independent of time. 

GHY term is calculated as 
\begin{equation}
I_{surf}^{future}=\frac{r^{z+d}L^d\Omega_d}{8\pi G}(\frac{t}2+r^*(\infty)-r^*(r))[(z+d)|f|+\frac{r}{2}\partial_{r}|f|]\Big|_{r=\epsilon_0}
\end{equation}
\begin{equation}
I_{surf}^{past}=\frac{r^{z+d}L^d\Omega_d}{8\pi G}(-\frac{t}2+r^*(\infty)-r^*(r))[(z+d)|f|+\frac{r}{2}\partial_{r}|f|]\Big|_{r=\epsilon_0}
\end{equation}
\begin{equation}
I_{surf}^{UV cutoff}=\frac{r^{z+d}L^d\Omega_d}{8\pi G}(2(r^*(\infty)-r^*(r))[(z+d)|f|+\frac{r}{2}\partial_{r}|f|]\Big|_{r=r_{max}}
\end{equation}
it can be seen that the total boundary terms is also independent of time. 

Joint terms from null boundaries insect with surface of past and future singularity vanish and joint terms from null boundaries insect with UV cutoff is independent of time \cite{Chapman:2016hwi}.So it has no contribution to complexity growth .

So we conclude that $\frac{dI}{dt}=0$ when $t<t_c$.

When $t>t_c$: 

the only difference is the intersections between the null boundaries with past singularity surface change to intersections between two null boundaries.
We calculate in the same way 
\begin{equation}
\label{feww}
I_{bulk}^1=-\int^{r_+}_{\epsilon_0}\frac{(z+d)L^d\Omega_d}{8\pi G}(\frac{t}2+r^*(\infty)-r^*(r))r^{z+d-1}dr
\end{equation}
\begin{equation}
I_{bulk}^2=-\int^{r_{max}}_{r_+}\frac{(z+d)L^d\Omega_d}{8\pi G}2(r^*(\infty)-r^*(r))r^{z+d-1}dr
\end{equation}
\begin{equation}
I_{bulk}^3=-\int^{r_+}_{r_m}\frac{(z+d)L^d\Omega_d}{8\pi G}(-\frac{t}2+r^*(\infty)-r^*(r))r^{z+d-1}dr
\end{equation}
and combine them all we get
\begin{equation}
I_{bulk}-I_{bulk}^0=-\int^{r_m}_0\frac{(z+d)L^d\Omega_d}{4\pi G}(\frac{t}2-r^*(\infty)+r^*(r))r^{z+d-1}dr
\label{delta1}
\end{equation}
where we have included the factor of two coming from two sides of WdW patch
For GHY boundary terms, in the absence of past surface at singularity, it also depends on time
\begin{equation}
\begin{aligned}
I_{surf}^{past}=\frac{r^{z+d}L^d\Omega_d}{8\pi G}(&-\frac{t}2+r^*(\infty)\\
&-r^*(r))[(z+d)|f|+\frac{r}{2} \partial_{r} |f|]\Big|_{r=\epsilon_0}
\end{aligned}
\end{equation}

\begin{equation}
\begin{aligned}
I_{surf}-I_{surf}^{0}&=-2I_{past} \\
&=-\frac{r^{z+d}L^d\Omega_d}{4\pi G}(-\frac{t}2+r^*(\infty)-r^*(r))\\
&\times[(z+d)|f|+\frac{r}{2}\partial_{r} |f|]\Big|_{r=\epsilon_0}
\label{delta2}
\end{aligned}
\end{equation}
where factor of two comes from the two sides of WdW patch.

Joint terms from null boundaries insect with surface of past and future singularity vanish and joint term from null boundaries insect with UV cutoff is independent of time.So we only need to consider the intersection of two past null boundaries at $r=r_{m}$,where $r_{m}$ satisfy equation:
\begin{equation}
\frac{t}{2}+r^*(r_m)-r^*(\infty)=0
\end{equation}
for $\delta t=t-t_c$,we can rewrite the above equation in 
\begin{equation}
\frac{\delta t}2=r^*(0)-r^*(r_{m})
\end{equation}
varying it with t,we get 
\begin{equation}
\frac{dr_m}{dt}=-\frac12 r^{z+1}_m f(r_m)
\label{drdt}
\end{equation}
We choose the normal vectors of null boundaries to be 
\begin{equation}
k_L=\alpha(dt+\frac{dr}{r^{z+1}f})
\end{equation}
\begin{equation}
k_R=-\alpha(dt-\frac{dr}{r^{z+1}f})
\end{equation}
The joint term is 
\begin{equation}
\begin{aligned}
I_{joint}=\frac1{8 \pi G}\oint a ds&=\frac1{8 \pi G}\oint\log(-\frac12{k_L}\cdot{k_R})ds\\
&=\frac{L^d\Omega_d r^d}{8\pi G}\log(\frac{-\alpha^2}{L^2 r^{2z} f})|_{r=r_m}
\label{delta3}
\end{aligned}
\end{equation}
combine equations ($\ref{delta1}$)($\ref{delta2}$)($\ref{drdt}$)($\ref{delta3}$),we can get 
\begin{equation}
\begin{split}
\frac{dI}{dt}&=\frac{dI_{bulk}}{dt}+\frac{dI_{surf}}{dt}+\frac{dI_{joint}}{dt}\\
&=-\frac{L^d\Omega_d}{8\pi G}r_m^{z+d}+\frac{(z+d)r_+^{z+d}L^d\Omega_d}{16\pi G}\\
&+\frac{L^d\Omega_d}{8\pi G}[-\frac{d}2 f(r_m) r_m^{z+d}\log(\frac{-\alpha^2}{L^2 r_m^{2z} f(r_m)})+z r_m^{d+z}+\frac{d-z}2 r_+^{z+d})]\\
&=\frac{L^d\Omega_d}{8\pi G}r_m^{z+d} (z-1)+\frac{d r_+^{z+d}L^d\Omega_d}{8\pi G}\\
&-\frac{d L^d\Omega_d}{16\pi G}[ f(r_m) r_m^{z+d}\log(\frac{-\alpha^2}{L^2 r_m^{2z} f(r_m)})]
\label{final}
\end{split}
\end{equation}
At the late time situation,$r_m\rightarrow r_+$ and $f(r_m)\rightarrow 0$,so the last term of (\ref{final}) is zero.Then the complexity growth can be simplify to
\begin{equation}
\frac{dI}{dt}=\frac{(d+z-1) r_+^{z+d}L^d\Omega_d}{8\pi G}
\end{equation}
which recover the late time result. 
To see whether the limit is approached from above or below, we plot the relation between complexity growth rate and time in four dimensions for various z in Fig4. 
\begin{figure}
\includegraphics[width=0.55\textwidth]{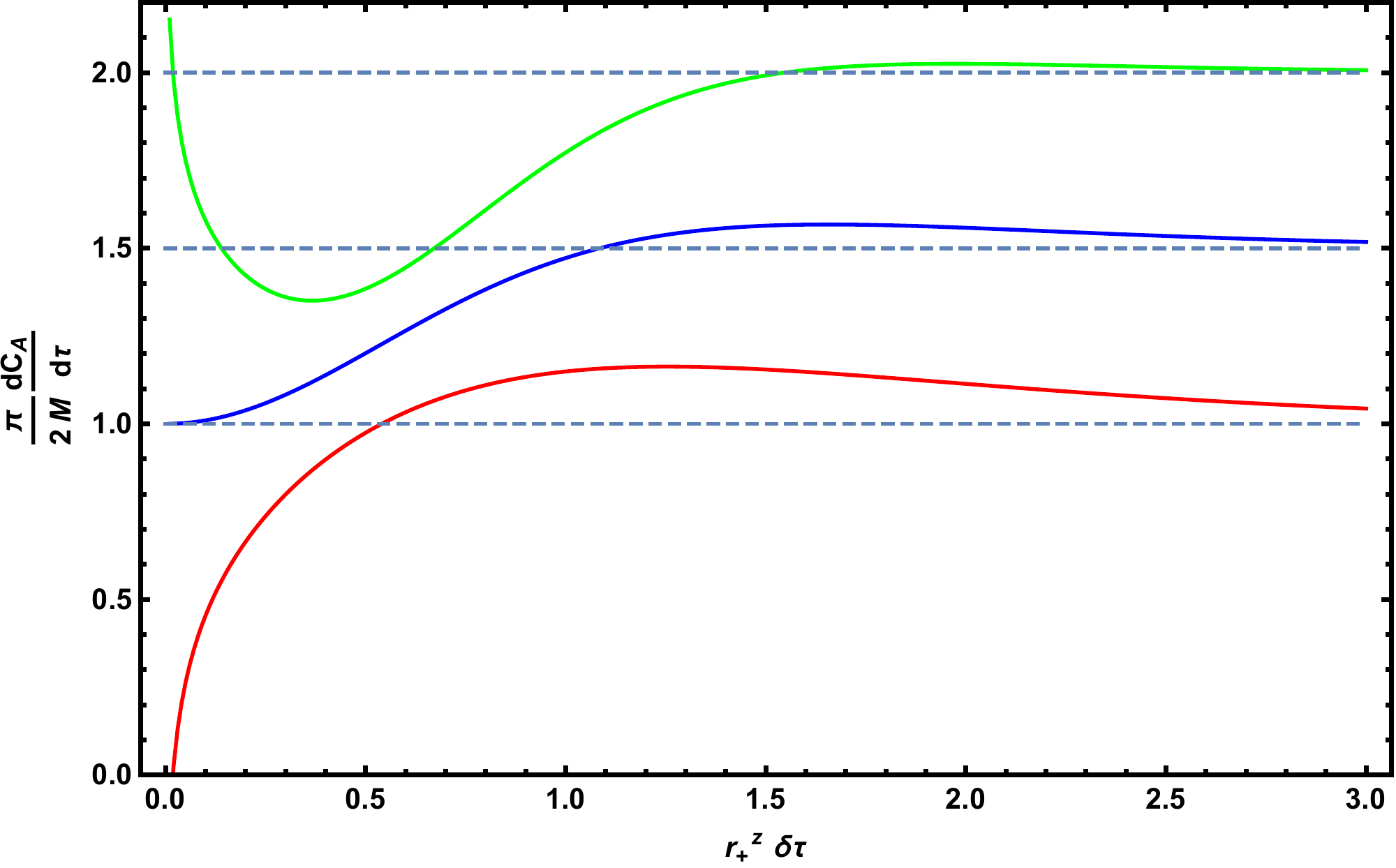}\\
  \caption{The relation between complexity growth rate and boundary time, we choose the normalization factor $\alpha=L r_{+}^{z}$, and green/ blue/red lines correspond respectively to $z=3$/$z=2$/$z=1$, we find it approaches the late time bound from above, and for z sufficiently large, the complexity growth rate experience a decreasing period during early time }
  \label{fig:complexity growth rate in different z}
\end{figure}
\section{Conclusion and discussion}
\label{Con}
In this paper, we investigate the effect of dilaton that is coupled to Maxwell field on the holographic complexity growth . In Section two, we investigate the black hole solution which is asymptotically AdS . In this case we find the presence of dilaton charge can slow down the computation rate, and when we take into account the full time evolution, we find the complexity growth rate approaches the late time bound from above like Ref \cite{Carmi:2017jqz}. In Section Three, we investigate the case which is asymptotically Lifshitz, in this case, dilaton is essential to support anisotropic vacuum structure.  We find that in this case, the Lloyd bound is violated even in the late time limit.  We investigate the full time evolution of the holographic complexity growth, it approaches the late time bound from above , moreover, for z sufficiently large , it exhibit an interesting behavior in time which is above critical time but much earlier than late time. 

Although the above two cases are both solution in EMD theory, the roles dilaton play are very different. For the AdS dilaton example, dilaton charge and electro charge are free parameters and appear in the black hole solutions. While in Lifshitz case, dilaton and maxwell field are added to maintain the anisotropic scaling background. It doesn't contribute to the black hole solution, which means that the black hole only has one free parameter M. Holographically, eternal AdS black hole is dual to thermofield double state $| TFD \rangle$, in Schwartzchild case, 
\begin{equation}
| TFD \rangle =\sum_{n} e^{-\beta E_{n}} | n_{L} \rangle |n_{R} \rangle
\end{equation}
In RN case 
\begin{equation}
| TFD \rangle =\sum_{n} e^{-\beta (E_{n}-\mu Q)}  | n_{L} \rangle |n_{R} \rangle
\end{equation}
what appears on the exponential depends on the thermodynamic relation. In Lifshitz case, because the charge is fixed cosntant. The first law of thermodynamics is just $dM=T dS$. So from boundary respect, there is no chemical potential term in thermofield double state. In other words, Ref.\cite{Brown:2015lvg} gives the complexity growth inequality
\begin{equation}
\frac{dC}{dt} \leqslant \int_{gs}^{S} T dS
\end{equation}
so from thermodynamical relation $dM=T dS$, we conclude that the complexity growth rate will only depend on the mass despite the presence of matter field in our Lifshitz case. 

Ref.\cite{Yang:2016awy} proves that the action growth obeys the bound 2M under the following two conditions : (1) the matter field locates outside the outmost killing horizon (2) the strong energy condition is obeyed. In our two cases, the matter field extends into the killing horizon,so it doesn't satisfy the condition required in \cite{Yang:2016awy}. In AdS dilaton case, the bound $2M$ is satisfied ,while in Lifshitz case, it is violated. So the requirements in \cite{Yang:2016awy} are too restrictive, our calculation shows that weaker condition is needed to prove the late time bound from the bulk side and distinguish these two cases. 

Recent study \cite{Swingle:2017zcd} investigate the action growth of the hyper-scaling violation background in EMD theory.They found that it depends on two parameters, the "dynamical exponent" $z$ and hyperscaling violation exponent $\theta$. They showed when $\theta=0$, their results match our Lifshitz result. 

There are several interesting directions to pursue, the reason of the violation of the Lloyd bound is still unclear.The EMD theory is not expected to be a UV complete theory of quantum gravity, so it would be interesting to take into account the stringy effect near the singularity and recalculate the action growth.Moreover, for full time evolution of complexity growth in Lifshitz black brane, for z sufficiently large, such as $z=3$, the complexity growth rate exhibits new behaviors that are not found in \cite{Carmi:2017jqz}, it is interesting to investigate what it means physically when z becomes large. \footnote{Recently this problem is investigated in Ref.\cite{Alishahiha:2018tep} by adding a counter term to remove the normalization ambiguity in the joint term. There is no strong motivation to add this term to remove the ambiguity unless field theory suggests us to do that. However as Ref. \cite{Jefferson:2017sdb} suggests , this normalization ambiguity is somehow related to the choice of reference state in the field theory construction of complexity}Although CA conjecture proposed in \cite{Brown:2015bva,Brown:2015lvg} passes many nontrivial tests including switchback effect \cite{Stanford:2014jda}, it is still unproved.Recent studies \cite{Carmi:2017jqz} implies that although CA conjecture captures some essence in complexity, it needs to be revised. Our work's result on Lifshitz black hole also gives the evidence that the revision is needed in CA duality. 

Apart from CA duality, there are also other proposals that relate complexity to other bulk quantities, such as \cite{Susskind:2014rva,Couch:2016exn,Alishahiha:2015rta,Momeni:2016ira,Momeni:2016ekm}, it is interesting to calculate complexity growth rate in these proposals and investigate if our results still hold. We will discuss the thermodynamic volume proposal \cite{Couch:2016exn} in the following

The Ref. \cite{Couch:2016exn} related the complexity on the boundary to the spacetime volume of WdW patch and gave the proposal complexity-volume duality 2.0. It is 
\begin{equation}
C \sim \frac{1}{\hbar} P (spacetime volume)
\end{equation}
In our Lifshitz case, because of following relation
\begin{equation}
\begin{aligned}
R-2\Lambda-\frac12 \partial_\mu\phi\partial^\mu\phi-\frac14 e^{\lambda\phi}F_{\mu\nu}F^{\mu\nu}=\frac{-2(z+d)}{L^2}
\end{aligned}
\end{equation}
the bulk action contribution is proportional to the spacetime volume 
\begin{equation}
I_{bulk}=-\frac{z+d}{8 \pi G L^{2}} \int d^{d+2} x \sqrt{-g} 
\end{equation}
so there is no essential difference between the volume and bulk action of WdW patch. 
We calculate the complexity growth rate using CV 2.0 proposal explicitly. 
In extended phase space, the pressure is identified as the cosmological constant
\begin{equation}
P=-\frac{\Lambda}{8\pi G}=\frac{(z+d)(z+d-1)}{16 \pi G L^{2}}
\end{equation}
The spacetime volume inside the WdW patch is calculated from \ref{feww} . And when $t>t_{c}$, the time dependence of the volume is 
\begin{equation}
\frac{dV}{dt}=\frac{L^{d+2}\Omega_{d} r_{m}^{z+d}}{z+d}
\end{equation}
After taking the late time limit ( $r_{m} \to r_{+}$). The complexity growth rate is 

\begin{equation}
\label{dcdt}
\frac{dC}{dt}=\frac{(z+d-1)M}{d}
\end{equation}
We see in the complexity-volume 2.0 proposal, the result is half of the CA proposal, the Lloyd bound 2M is satisfied when z is not large enough, but as in Ref.\cite{Couch:2016exn}  the Lloyd bound can be altered up to a pre-factor
\begin{equation}
\frac{dC}{dt} \leqslant  \frac{\alpha E}{ \pi \hbar}
\end{equation}
the Ref.\cite{Couch:2016exn} shows that in AdS-RN case, this bound is satisfied for $\alpha=1$. In CV 2.0 proposal if we set $\alpha=1$ to be the Lloyd bound, this bound is violated in the Lifshitz case at late time because of the anisotropic scaling z as in the CA proposal.  

Moreover, as conjectured in Ref. \cite{Couch:2016exn}, in various cases ,the late time value of the time derivative of spacetime volume of the WdW patch reduces to the thermodynamical volume( or the difference between two thermodynamic volume in the case of two horizons). In Lifshitz case, this late time limit of spacetime volume derivative reads 
\begin{equation}
V=\frac{L^{d+2}\Omega_{d} r_{+}^{z+d}}{z+d}
\end{equation}
when $z=1$ in four dimension, it reduces to the familiar thermodynamic volume of Schwartzchild case $\frac{4 \pi r^{3}_{+}}{3}$. While when $z \neq 1$, because the profile of matter field depends on the cosmological constant nontrivially, it is hard to use Iyer-Wald formalism \cite{Iyer:1994ys} to derive the explicit expression of the thermodynamic volume. So in the further work , it will be interesting to derive the thermodynamic volume in Lifshitz case and see if it is the same as our late time limit result.  It will provide a convincing evidence for the conjecture in Ref.\cite{Couch:2016exn} that thermodynamic volume is essential for complexity growth.

Although there have been many discussions on the complexity in holographic side, a concrete definition of complexity in quantum field theory still lacks. Recently, there are many works concerning this question.  \cite{Jefferson:2017sdb,Chapman:2017rqy} propose some definitions about complexity in field theory side using the Finsler geometry and Fubini-Study metric. \cite{Caputa:2017yrh} discretizes the Euclidean path integral and defines the "path integral complexity" in terms of Liouville action. For concrete calculation in field theory side, see also\cite{Kim:2017qrq,Yang:2017nfn, Hashimoto:2017fga}. 
Once a concrete and calculable definition of complexity is given, the problem of CA duality conjecture will be clarified to large extent. Moreover, apart from the problems from CA conjecture, the assumptions made in the process to derive the Lloyd bound should also be checked \cite{Cottrell:2017ayj}.

\begin{acknowledgments}
We want to thank Prof. Ronggen Cai for useful advice and encouragement during this work . And we also want to thank Run-Qiu Yang, Shao-Jiang Wang for useful help and discussions during this work. 
\end{acknowledgments}

\bibliographystyle{utphys}
\bibliography{ref}

\end{document}